\documentclass[a4paper]{amsart}

\usepackage{amsmath}  
\usepackage{amsfonts}
\usepackage{times}
\usepackage{helvet}
\usepackage[numbers]{natbib} 
\usepackage{url}

\usepackage{array}
\usepackage{graphicx}
\usepackage{color}
\usepackage{tikz}
\usepackage[swedish,english]{babel}
\usepackage[latin1]{inputenc}
\usepackage[T1]{fontenc}

\newcommand{\R}{\mathbb{R}}

\newcommand{\Z}{\mathbb{Z}}
\newcommand{\LL}{{\mathcal{L}}}
\newcommand{\EE}{\mathbb{E}}
\newcommand{\Ecal}{\mathcal{E}}
\definecolor{ogreen}{rgb}{0.3194,0.6223,0.0766}    

\newcommand{\omegagammaepsilon}{\Omega_{\Gamma_{\varepsilon}}}

\title{Free path lengths in quasi crystals}
\author{Bernt Wennberg}
\address{Department of mathematical sciences, Chalmers University of Technology, SE41296 Gothenburg, Sweden}
\address{Department of mathematical sciences, University of Gothenburg, SE41296 Gothenburg, Sweden}
\email{wennberg@chalmers.se}
\begin{document}

\maketitle

\begin{abstract}

{\small

 The Lorentz gas is a model for  a cloud of point
particles (electrons) in a distribution of scatterers in space. The
scatterers are often assumed to be spherical with a fixed diameter
$d$, and the point particles move with constant velocity between the
scatterers, and are specularly reflected when hitting a
scatterer. There is no interaction between point particles. An
interesting question concerns the distribution of free path lengths,
{\em i.e.} the distance a point particle moves between the scattering
events, and how this distribution scales with scatterer diameter,
scatterer density and the distribution of the scatterers. It is by now
well known that in the so-called Boltzmann-Grad limit, a Poisson
distribution of scatters leads to an exponential distribution of free
path lengths, whereas if the scatterer distribution is periodic, the
distribution of free path behaves asymptotically like a Cauchy
distribution. 

This paper considers the case when the scatters are distributed
on a quasi crystal, {\em i.e.} non periodically, but with a long range
order. Simulations of  a  one-dimensional model are presented, showing
that  the quasi crystal behaves very much like a periodic crystal, and
in particular, the distribution of free path lengths is not exponential.
}

\end{abstract}

\bigskip

\noindent
\section{Introduction}
\label{sec:introduction}

The Lorentz gas is a mathematical model for the motion of (point)
particles in e.g. a crystal, consisting of spherical, elastic
scatterers of radius $a$ with centers at a fixed set of points
$\Gamma_{\varepsilon}\subset\R^n$. A point particle moves in straight
lines between the obstacles, on which it is specularly reflected. 
At least two very different  scatterer 
distributions, $\Gamma_{\varepsilon}$ have been studied thoroughly:
the standard lattice $ \LL\subset\R^n$, with interstitial distance
$\varepsilon^{(n-1)/n}$, or a random distribution, where $\Gamma_{\varepsilon}$
is Poisson distributed with intensity $\varepsilon^{-(n-1)}$. In this
paper we are mainly concerned with the so called Boltzmann-Grad limit
of this system, and that corresponds to setting the scatter radius
$a=\varepsilon$ and letting $\varepsilon\rightarrow 0$. Very loosely
speaking, in this scaling the mean free path of the point particles
remain of order one as $\varepsilon\rightarrow 0$. Consider now a
density $f_{\varepsilon}(x,v,t]$ of point particles moving between the
scatterers. Gallavotti~\cite{Gallavotti1972} (see
also~\cite{Gallavotti1999book}) considered the random
obstacle distribution, and proved  that when $\varepsilon\rightarrow
0$, $f_{\varepsilon}(x,v,t)$ converges to a function that solves a
linear Boltzmann equation. On the other hand, Golse~\cite{Golse2008}
proved, based on results in~\cite{BourgainGolseWennberg1998}
and~\cite{GolseWennberg2000}, that in the periodic case, the limiting
particle distribution does not satisfy a linear Boltzmann
equation. Caglioti and Golse~\cite{CagliotiGolse2003,CagliotiGolse2008,CagliotiGolse2010} and Marklof and
Strömbergsson~\cite{MarklofStrombergsson2008,MarklofStrombergsson2010,MarklofStrombergsson2011,MarklofStrombergsson2011b}
independently found sharp estimates of the distribution of free path
lengths and that the Boltzmann-Grad limit corresponds to a  
Boltzmann like equation in an extended phases space that does describe
the evolution of a particle density in the limit of small
$\varepsilon$; whereas the results
in~\cite{CagliotiGolse2003,CagliotiGolse2008,CagliotiGolse2010} are
restricted to two space dimensions and rely on an independence
assumption, \cite{MarklofStrombergsson2008,MarklofStrombergsson2010,MarklofStrombergsson2011,MarklofStrombergsson2011b} provides a complete proof valid for any space
dimension. Related results valid for in the two-dimensional case, can
be found in~\cite{BocaGologanZaharescu2003, BocaZaharescu2007}, and
in~\cite{Bykovski2008}. Situations where scatterers are randomly place on a
periodic lattice have been considered in ~\cite{CagliotiPulvirentiRicci2000} and
~\cite{RicciWennberg2004}. 

Here we are interested in the case where the obstacles are distributed
as the atoms in a quasi crystal. By definition {\em a crystal} is ''a solid
with an essentially discrete diffraction pattern''~\cite{Senechal1995book}. It
has been known for a long time that periodic crystals in three
dimensions must belong to one of fourteen symmetry classes. A {\em
  quasi crystal} is a crystal whose diffraction pattern exhibits a
forbidden symmetry. The first reports on experimental results
indicating that such solids  exist were treated with suspicion by
the scientific community,  but in 2011 their discovery was awarded the Nobel Prize in Chemistry~\cite{NobelPrize2011Chemistry}.There are also many mathematical abstract constructions
that give the same results, starting {\em e.g.} from the Penrose
tiling~(see {\em e.g.}~\cite{Senechal1995book}). It
is then a natural question to ask whether 
the Boltzmann-Grad limit of a Lorentz gas in a quasi crystal behaves
more like the random or periodic case. In this paper we present
simulation results on a one-dimensional model, which give strong
support for the latter: the free path length distribution decays polynomially,
just as in the periodic case, whereas in the random case, the path
length distribution is exponentially decaying.

The next section gives precise definitions of the Lorentz model, and
of its Boltzmann-Grad limit, and some results concerning the periodic
and random cases are discussed. Thereafter, in Section~\ref{sec:quasi},
a construction of quasi crystals is given; that section is mainly
based on~\cite{Senechal1995book}.

Section~\ref{sec:simulation} presents the simulation method and the
results, and the paper ends with conclusions and some prospects for
future research in Section~\ref{sec:conclusions}.

\section{Free path length distributions in the Lorentz model}
\label{sec:lorentzmodel}

The discussion in this section is restricted to the Lorentz gas in two
dimensions, and hence we consider a point distribution
$\Gamma_{\varepsilon} \subset \R^2$, or more precisely, a family of
point distributions parametrized by $\varepsilon$. At each point
$p\in\Gamma_{\varepsilon}$ we put a circular obstacle with radius
$\varepsilon$ and center at $p$, and then we study the motion of a
point particle moving with constant speed, $|\mathbf{v}|=1$, along
straight lines between the obstacles, and 
specularly reflected when hitting an obstacle. Hence the phase space
for one point particle is $ \omegagammaepsilon\times S^1$,
where 
\begin{equation*}
  \omegagammaepsilon= \R^2 \setminus
\bigcup_{p\in\Gamma} \bar{B}_{p}(\varepsilon)
\end{equation*}
and
 $\bar{B}_{p}(\varepsilon)$ is the
{\em closed} ball of radius $\varepsilon$ centered at $p\in\R^2$. For
any initial point $(x_0,v_0)$ we define 
\begin{equation*}
  T_{\Gamma_{\varepsilon}}^t(x_0,v_0) = (x(t),v(t))\,,
\end{equation*}
the position of the point particle at time $t$, taking into account
all reflections on the set of obstacles. As long as the obstacles do
not overlap, this is well defined for all $t\ge 0$. Moreover, the 
the {\em free path length} is defined as 
\begin{equation*}
  \tau_{\varepsilon}(x_0,v_0)=\inf \left\{ t>0\,\Big| \, x_0+t v_0 \notin
    \omegagammaepsilon\right\} \,.
\end{equation*}
We also define the {\em path length distribution} 
\begin{equation}
\label{eq:pld}
  \phi_{\varepsilon}(]a,b[)= \lim_{R\rightarrow\infty} \frac{m(\{ (x,v)\in
    \left( \omegagammaepsilon\cap B_0(R)\right)\times S^1 \,|\, \tau_{\varepsilon}(x,v)\in
    ]a,b[)\}}{m( \omegagammaepsilon\cap B_0(R)) }\,.
\end{equation}
Here $m(\cdot)$ is the Lebesgue measure restricted to
$\omegagammaepsilon\times S^1$. In cases where, at least formally, one
can prove that for all intervals $[a,b]$, $\phi_{\varepsilon}(]a,b[)$
remains bounded from above and below when $\varepsilon\rightarrow 0$
one speaks of a {\em  Boltzmann-Grad limit}.

The two typical examples of point distributions
$\Gamma_{\varepsilon}$ considered in connection with the Lorentz gas
are the periodic  distributions $\Gamma_{\varepsilon}=  a(\varepsilon)
\Z^2$, and a Poissonean random distribution with intensity $a(\varepsilon)$. 

For a periodic distribution it is more natural to define the free path
length distribution by restricting to a lattice unit cell rather than
to a ball of radius $R$ as in~(\ref{eq:pld}). Of course, in the limit
$R\rightarrow\infty$ the result is the same. In this case  the
Boltzmann-Grad limit corresponds to choosing
$a(\varepsilon)=\varepsilon^{1/2}$. The path length distribution
$\phi_{\varepsilon}$ in the Boltzmann-Grad limit  has been studied 
in~\cite{BourgainGolseWennberg1998,GolseWennberg2000}, and then, with
very sharp bounds 
in~\cite{CagliotiGolse2003,CagliotiGolse2008,CagliotiGolse2010} using
the theory of continued 
fractions, and~\cite{MarklofStrombergsson2008,MarklofStrombergsson2010,MarklofStrombergsson2011,MarklofStrombergsson2011b} using methods based on
Ratner's theorem. The result is that $\phi_{\varepsilon}([T,\infty[)
\rightarrow C\,  T^{-1} $ asymptotically for large $T$ when
$\varepsilon\rightarrow 0$.

The distribution $\Gamma_{\varepsilon}$ is a Poisson distribution with
intensity $a(\varepsilon)$ if and only if for any set $A\subset \R^2$,
\begin{equation*}
  {\mathrm{Prob}}\left( \mbox{ no of points in } A = k \right) =
 \frac{ \left(m(A)a(\varepsilon)\right)^k}{k!}e^{-m(A)a(\varepsilon)}\,,
\end{equation*}
and in also, for $A\subset\R^2$ and $B\subset\R^2$ with $A\bigcup
B=\emptyset$, the number of points in  $\Gamma_{\varepsilon} \bigcup
A$ and  $\Gamma_{\varepsilon} \bigcup B$ are independent random
variables. Here the Boltzmann-Grad limit is achieved by setting
$a(\varepsilon)=\varepsilon^{-1}$, and letting $\varepsilon$ go to
zero. This distribution differs from the periodic one in several
fundamental aspects. First, obstacles may overlap, and hence the
trajectories $(x(t),v(t))$ cannot always be continued
uniquely. However, the measure of such, bad trajectories goes to zero
when $\varepsilon\rightarrow 0$. Secondly, before taking the limit
$\R\rightarrow \infty$ in the definition of the path length
distribution, the expression in the right hand side of
equation~(\ref{eq:pld}) is a random variable, but one that converges
to a deterministic value both  when
$R\rightarrow\infty$, and when $\varepsilon\rightarrow 0$. The path
length distribution $\phi_{\varepsilon}([T,\infty[)$ converges to a
distribution $\phi([T,\infty[)\sim \exp(-cT)$. The Lorentz gas with
Poisson distributed obstacles has been studied in detail by
Gallavotti~\cite{Gallavotti1972}, who proved that if $f_{\varepsilon,0}(x,v)\in
L^1\left(\omegagammaepsilon\times S^1\right)$  are densities of
initial points for point particles, and $f_{\varepsilon}(x,v,t) =
\EE\left[f_{\varepsilon,0}(T^{-t}_{\Gamma_{\varepsilon}}(x,v))\right]$,
then, assuming that $f_{\varepsilon,0}\rightarrow f_0\in
L^1(\R^2\times S^1)$, it follows that  $f_{\varepsilon}(x,v,t) \rightarrow
f(x,v,t)$, which solves a linear Boltzmann equation.

A different class of random distributions of scatterers can be
constructed starting 
from a periodic distribution, by removing obstacles randomly,
independently,  with some probability $p(\varepsilon)$. Although for a
given, positive $\varepsilon$, the behavior is rather different from
the Poissonean case, this difference disappears in the limit as
$\varepsilon\rightarrow 0$, and in particular, it is possible to
rigorously derive a linear Boltzmann equation starting from such
distributions (see~\cite{CagliotiPulvirentiRicci2000,
  RicciWennberg2004}). 

\begin{figure}
  \centering
\begin{tikzpicture}[scale=3.0]
\draw[thin] (-0.5,-0.5) -- (-0.5,0.5) -- (2.5,0.5) -- (2.5,-0.5)
--(-0.5,-0.5);
\draw[thin] (0.5,-0.5) -- (0.5,0.5);
\draw[thin] (1.5,-0.5) -- (1.5,0.5);
\draw (-0.5,0) -- (2.5,0);
\draw[blue,fill] (0.0,0.0) circle (0.05cm);
\draw[blue,fill] (1.0,0.0) circle (0.05cm);
\draw[blue,fill] (2.0,0.0) circle (0.05cm);
\draw[red,thick] (-0.1,0)  node[below] {\color{black} $ q_0$} --
(0.265,0.5);
\draw[thin, dashed] (-0.1,0) -- (-0.1,0.5);
\draw[thin] (0.076,0.177) node[anchor=south east] {$\theta$} arc (45:100:0.25cm);
\draw[red,dashed] (0.265,0.5) -- (0.265,-0.5);
\draw (0.66,0.0) node[below]{$q_1$};
\draw[red,thick] (0.265,-0.5) -- (0.995,0.5);
\draw[red,dashed] (0.995,0.5)--(0.995,-0.5);
\draw (1.4,0.0) node[below]{$q_2$};
\draw[red,thick] (0.995,-0.5)--(1.725,0.5);
\draw[red,dashed] (1.725,0.5) --(1.725,-0.5);
\draw (2.1,0.0) node[below]{$ q_3$};
\draw[red,thick] (1.725,-0.5) -- (2.455,0.5);

\end{tikzpicture}
\caption{Computing the free path length in a periodic Lorentz gas is
  equivalent to to a discrete map.}
\label{fig:per1}
\end{figure}
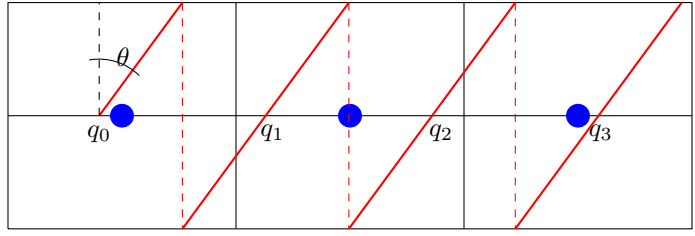

The following section gives examples of mathematical constructions of
quasi crystals, and the corresponding Lorentz gas. However, computing
long point particle trajectories in a quasi crystal Lorentz gas
is very time consuming, and therefore the 
simulation results presented in this paper  are performed on a
one-dimensional discrete model. In the two-dimensional periodic case,
the path length distribution can be computed almost exactly by
analyzing the  discrete map given by the
consecutive intersections of a trajectory with lines parallel to the
lattice containing lattice points, as in Figure~\ref{fig:per1}. We
assume here that $a=1$. For an
initial point $(x_0,v)$ with $x_0=(q_0,\tilde{q}_0)$ sitting on a
horizontal line as in 
the figure, and with $v\in S^2$ having an angle $\theta$ to the
vertical line, the distance between two consecutive points is
$\tan(\theta)$. To find the free path length of a trajectory in
a periodic Lorentz gas is then (almost) equivalent to computing 
\begin{equation}
\label{eq:diskfreepath}
  k=\min \left\{ j\; \big| \; \mbox{dist}(q_j, \Z) \le
    \sqrt{\varepsilon}/2\right\}. 
\end{equation}
In fact, we will have $\tau_{\varepsilon}(x_0,v) = \sqrt{\varepsilon}
(k\pm 1) /\cos\theta$. 

\section{Quasi crystals}
\label{sec:quasi}

The material in this section is mostly taken from M. Senechal's book
''Quasi Crystals and Geometry''~\cite{Senechal1995book}, which gives a
good introduction to the subject. 

There are several ways of construction point sets in $\R^n$ that
satisfy the requirement for being a quasi crystal: ''an essentially
discrete diffraction pattern exhibiting a forbidden symmetry''. One of
these is the so called projection method: 

  \begin{itemize}
  \item Let $\LL^p$ be a point lattice in $ \R^k$  (usually the
    standard lattice), and let $K$ be the  unit cell in the lattice
    that contains the origin.
\item Let  $ \Ecal$ be an $ n$-dimensional subspace of $ \R^k$, such that $
  \Ecal \cap \LL^p = \{ 0 \}$, and let  $ \Ecal^{\perp}$be  the orthogonal
  complement of $\Ecal$. 
\item Let $\Pi$ and $\Pi^{\perp}$ be the orthogonal
  projections on $\Ecal$ and $\Ecal^{\perp}$ respectively.
\item Let $  X = \LL^p \cap \left(  \Pi^{\perp}(K) \oplus
    \Ecal\right)$, {\em i.e.} the set of lattice points contained in the
  strip (or cylinder) obtained by translating the unit cell $K$ along $\Ecal$.
  \end{itemize}

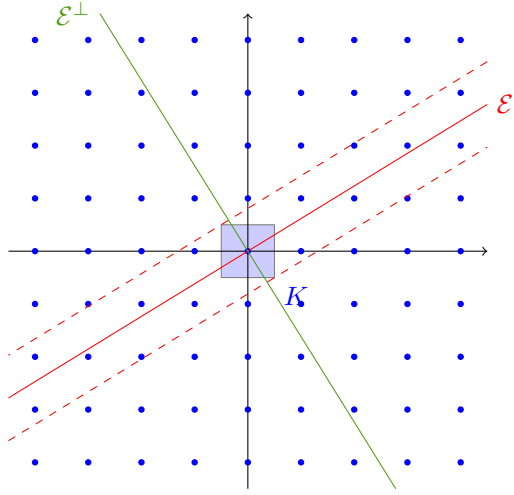
\begin{figure}
    \centering

\begin{tikzpicture}[scale=0.7]
\filldraw[fill=blue!20!white,draw=gray] (-0.5,-0.5) -- (0.5,-0.5) -- 
(0.5,0.5) -- (-0.5,0.5) -- cycle;
\draw (0.5,-0.5) node[anchor=north west] {$\color{blue} K$};
\draw[thin,->] (-4.5,0) -- (4.5,0); 
\draw[thin,->] (0,-4.5) -- (0,4.5); 
\foreach \x in { -4,-3,...,4}
\foreach \y in { -4,-3,...,4}
\draw[blue, fill] (\x,\y) circle (0.05cm);
\draw[red] (-4.5,-2.78) -- (4.5,2.78) node[right]{$\Ecal$};
\draw[ogreen] (2.78,-4.5) -- (-2.78,4.5) node[left]{$  \Ecal^{\perp}$};
\draw[red, dashed] (4.5,3.59) -- (-4.5,-1.97);
\draw[red,dashed] (-4.5,-3.59) -- (4.5,1.97);
\end{tikzpicture}
    \caption{A two dimensional representation of the projection method for constructing quasi crystals}
    \label{fig:quasi1}
  \end{figure}

The quasi crystal is $\Lambda= \Pi(X)$, the orthogonal projections of $X$ on
$\Ecal$. This set is discrete ($\inf \left\{|p_1-p_2|\; \big|\; p_1,p_2\in
  \Lambda\right\} =r_0>0$), non periodic and relatively dense
(meaning that there is $R_0>0$ such that every sphere in $\Ecal$ of
radius greater than  $R_0$ contains at least one point of $\Lambda$)
(see Figure~\ref{fig:quasi1}). That  $\Lambda$ is discrete and
relatively dense implies in particular that it is possible to define a
scaling that corresponds to the Boltzmann-Grad limit.

The theory of quasi crystals started with the (experimental)
discovery of a material exhibiting a five-fold
symmetry. Mathematically this can be constructed using the projection
method, starting from the regular lattice in $\R^5$. A unit cell in
this lattice is the hyper cube $Q_5$ with 32 vertices
\begin{equation*}
  \frac{1}{2}\sum_{j=1}^{5} \alpha_j e_j, \qquad,\alpha_j \in\{-1,1\}\,
\end{equation*}
where $e_1=(1,0,0,0,0), ...., e_5=(0,0,0,0,1)$. A five fold symmetry
is given by cyclic permutation of the unit vectors $e_j$, a
transformation that can be represented by the rotation matrix
\begin{equation*}
  \mathcal{A}= \left[
\begin{array}{ccccc}
0&0&0&0&1\\
1&0&0&0&0\\
0&1&0&0&0\\
0&0&1&0&0\\
0&0&0&1&0
\end{array}
\right]
\end{equation*}
which by an orthogonal change of variables becomes
\begin{equation*}
   \left[
\begin{array}{ccccc}
\cos(\frac{2\pi}{5})&-\sin(\frac{2\pi}{5}) &0&0&0\\
\sin(\frac{2\pi}{5})&\cos(\frac{2\pi}{5})&0&0&0\\
0&0&\cos(\frac{4\pi}{5})&-\sin(\frac{4\pi}{5})&0\\
0&0&\sin(\frac{4\pi}{5})&\cos(\frac{4\pi}{5})&0\\
0&0&0&0&1
\end{array}
\right]
\end{equation*}
In these new coordinates, $\Ecal$ will be the subspace spanned by
$(1,0,0,0,0)$ and $(0,1,0,0,0)$. The quasi crystal will then be the
the set
\begin{equation*}
  \Pi\left(\LL^p \cup   \left( \Pi^{\perp}(\mathcal{Q}_5) \oplus
    \Ecal\right)\right)\subset \Ecal\,.
\end{equation*}

A one-dimensional quasi crystal can be defined in the similarly as a
projection on a one dimensional subspace of $\R^2$. Particular
examples, which are convenient for numerical computations, are the
point  sequences known as Fibonacci sequences.  They  are defined by taking 
\begin{equation*}
  \Ecal = \left\{ x\in \R^2\,|\, x\cdot \omega =0\right\}\,,
\end{equation*}
where
\begin{equation*}
 \omega = (-1,\tau) \quad \mbox{with}\quad  \tau=\frac{1+\sqrt{5}}{2}\,.
\end{equation*}
Let $\nu = |\omega| = \sqrt{1+\tau^2}$. There is an explicit formula
for computing the sequence of points that constitute the
one-dimensional quasi crystal $\mathcal{F}$ obtained by this construction:
\begin{equation}
\label{eq:fiboseq}
  \mathcal{F} = \bigcup_{m\in\Z} \{ x_m\} \quad \mbox{with } \quad
  x_m = \frac{m}{\nu} + \frac{1}{\tau \nu} \left\| \frac{m}{\tau}\right\|\,,
\end{equation}
where $\|\cdot\|$ denotes the distance to the
nearest integer. This sequence has many interesting properties, and
in particular it satisfies the criteria that defines a quasi
crystal. A particularity is that the interval between two consecutive
points are of exactly two kinds, short and long:
\begin{equation*}
  x_m-x_{m-1} = \frac{1}{\nu} \quad \mbox{or} \quad \frac{1}{\nu} +
  \frac{1}{\nu \tau}\,.
\end{equation*}

As we have seen above, there is a very direct connection between the
periodic Lorentz gas in two dimensions and a discrete model, at least
when it comes to computing the free path-length distribution,
illustrated in Figure~\ref{fig:per1} and
Equation~(\ref{eq:diskfreepath}). Here we define the rescaled free path length
\begin{equation}
\label{eq:fibofreepath}
  \tau_{\varepsilon}(q_0,v) = \varepsilon \min \left\{ j\;  \big| \;
    \mbox{dist}(q_0+j v, \mathcal{F}) \le \varepsilon/2\right\}. 
\end{equation}
(Note that compared to Equation~(\ref{eq:diskfreepath}) the scaling factor
$\sqrt{\varepsilon}$ is replaced by $\varepsilon$, the only reason
being to make notation more clean).

In the following section we simulate trajectories in order to  estimate the
path length distribution and compare with simulation results when
$\mathcal{F}$ is replaced by $\alpha^{-1} \Z$, and with a Poisson stream of
points $x_j$ with intensity $\alpha$. The factor $\alpha$ is chosen so
that all three point distributions have the same density, asymptotically:
\begin{equation*}
 \alpha =  \lim_{R\rightarrow\infty} \frac{\#( [-R,R] \bigcap \mathcal{F})}{2R}
  = \frac{\tau^2}{\nu}\,.
\end{equation*}
This is not equivalent to the free path length distribution in a
two-dimensional Lorentz gas as it is in the periodic case, however.

\section{Simulation method and results}
\label{sec:simulation}

Here we study the distribution of free path lengths of a point
particle jumping on the real line, {\em i.e.} the number of jumps
needed before the particle falls into an interval of width
$\varepsilon$ with center at either the quasi crystal $\mathcal{F}$,
as defined previously, the periodic lattice $\frac{\nu}{\tau^2} \Z$,
or on a Poisson distributed set of points, $\Gamma_{\varepsilon}$ with
intensity $\tau^2/\nu$.  The intensity of the Poisson distribution and
the scaling of the periodic lattice are chosen in order that all three
obstacle distributions have the same density.

The random starting points $q_0$ are chosen randomly, uniformly in the
interval \\ $10 000 \nu/\tau^2$. Obviously, for computing the path length
distribution in the random obstacle distribution this is not
necessary, $q_0=0$ would give exactly the same result. Moreover, in
this case the distribution can be computed analytically in the limit
of small $\varepsilon$. That case is included in the simulation only
for comparison. In the periodic distribution of scatterers it would be
more natural to chose a random initial point uniformly in the interval
$[0,\nu/\tau^2]$, but by definition the quasi crystal is not periodic,
and therefore it is natural to chose a larger interval. 

For the simulations presented here, $v$ is chosen uniformly in
$[0,\nu/\tau^2]$. To obtain a result completely equivalent to the
periodic Lorentz gas in two dimensions, one should have taken
$v=\frac{\nu}{\tau^2}\tan\theta$ with $\theta$ uniformly chosen in the
interval $[0,\pi/4]$, but simulations with different jump length
distributions show give very similar results, and this is not
presented here. However, we have only tried distributions with
bounded densities.

The position of the point particle after  $j$ jumps is denoted
$q_j=q_0+v j$, and the points $x_m\in \mathcal{F}$ in the Fibonacci
sequence are computed with the formula~(\ref{eq:fiboseq}). The points
in the Poisson distribution are computed independently for each
trajectory $q_m$.

The simulation procedure is then
\begin{enumerate}
\item Chose $q_0$ and $v$ randomly
\item Compute $k_{\mathcal{F},\varepsilon} = \min \left\{ j\;  \big| \;
    \mbox{dist}(q_0+j v, \mathcal{F}) \le \varepsilon/2\right\}$
\item Compute $k_{\mathcal{\Z},\varepsilon} = \min \left\{ j\;  \big| \;
    \mbox{dist}(q_0+j v, \frac{\nu}{\tau^2}\Z) \le
    \varepsilon/2\right\}$
\item Compute $k_{\Gamma,\varepsilon} = \min \left\{ j\;  \big| \;
    \mbox{dist}(q_0+j v, \Gamma) \le
    \varepsilon/2\right\}$, where $\Gamma$ is the random obstacle
  distribution. Note that for a given trajectory $q_k$, the free path
  time is a random variable, depending on the realization of the
  obstacle distribution. Only one realization of the random obstacle
  has been chosen. 
\item Make a record of each of these path lengths, and repeat a large
  number $N$ times.
\end{enumerate}

 For all simulation results presented below, $N=2\times 10^7$, and the results 
 results have been used to estimate $\mathrm{Prob}[ k_{*,\varepsilon} \ge
  K]$, where $*$ represents the different obstacle distributions. In
  connection with the Lorentz equation, the relevant quantity is the
  path length distribution expressed in time units,
$\mathrm{Prob}[ \varepsilon k_{*,\varepsilon} \ge T]$, which is
estimated as
\begin{equation}
\label{eq:pathlenghts2}
  \phi_{*\varepsilon}([T,\infty[)= \frac{\mbox{Number of trajectories with
    }k_{*,\varepsilon}\ge T \, \varepsilon^{-1}}{N} 
\end{equation}

\begin{figure}
  \centering
    \includegraphics[width=0.8\textwidth]{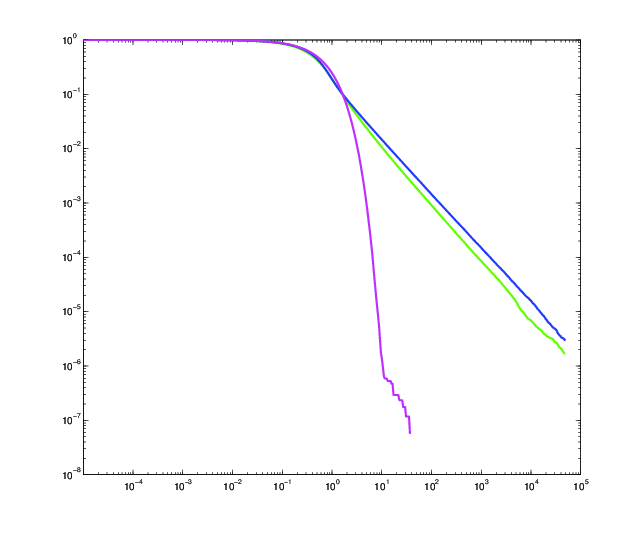}    
  \caption{The free path length distribution based on $10^7$
    trajectories and $\varepsilon=10^{-5}$. The curves show the result
    for the Fibonacci distribution of scatters (green), the periodic
    distribution (blue) and the random distribution (magenta).} 
  \label{fig:sim1}
\end{figure}

The first example, shown in Figure~\ref{fig:sim1}, compares the free
path length distributions with the three different distribution of
scatterers that we consider here for $\varepsilon=10^{-5}$. The graph,
plotted in logarithmic scale, show the $1/T$-behavior of
$\phi_{*,\varepsilon}([T,\infty[)$ both for the quasi crystal
$\mathcal{F}$ and for the periodic distribution, while the random
distribution gives a different result (an exponential decay, as expected).  

\begin{figure}
  \centering
    \includegraphics[height=0.5\textwidth]{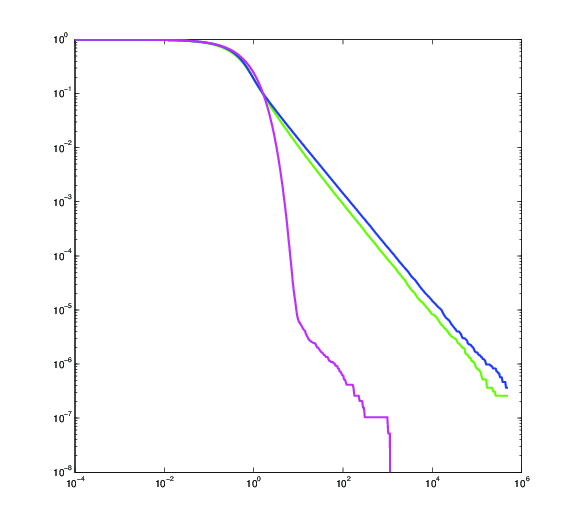}~%
\includegraphics[height=0.5\textwidth]{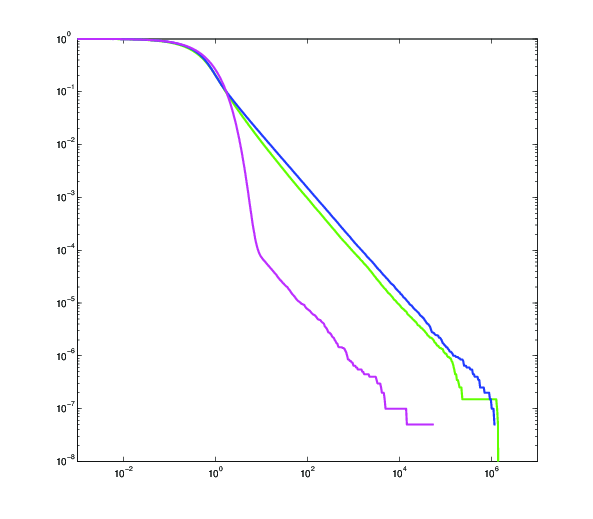}        
  \caption{The free path length distribution computed as in
    Figure~\ref{fig:sim1}  but with $\varepsilon=10^{-4}$ (left) and
    $\varepsilon=10^{-3}$ (right)} 
  \label{fig:sim2}
\end{figure}

Figure~\ref{fig:sim2} shows the same as Figure~\ref{fig:sim1} but with
$\varepsilon=10^{-4}$ (left) and $\varepsilon=10^{-3}$ (right). Note
that the path length distribution for the random obstacle distribution
is exponentially  only in  of the exponential distribution for large
$T$. This corresponds the fraction of trajectories with a jump length
$v<\varepsilon$ which are stopped by the first obstacle they reach.

\begin{figure}
  \centering
    \includegraphics[height=0.5\textwidth]{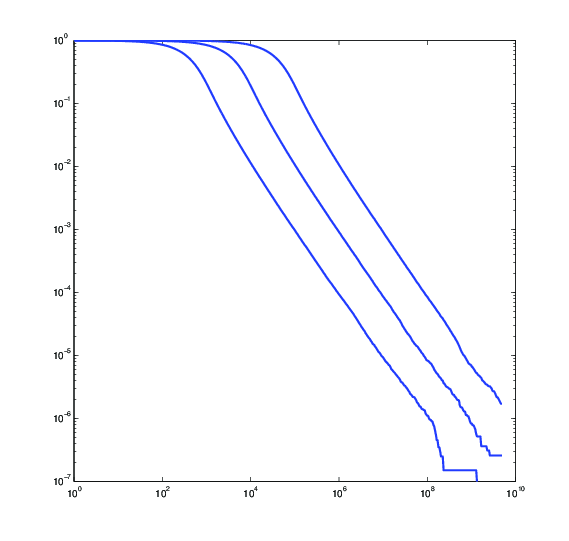}~%
\includegraphics[height=0.5\textwidth]{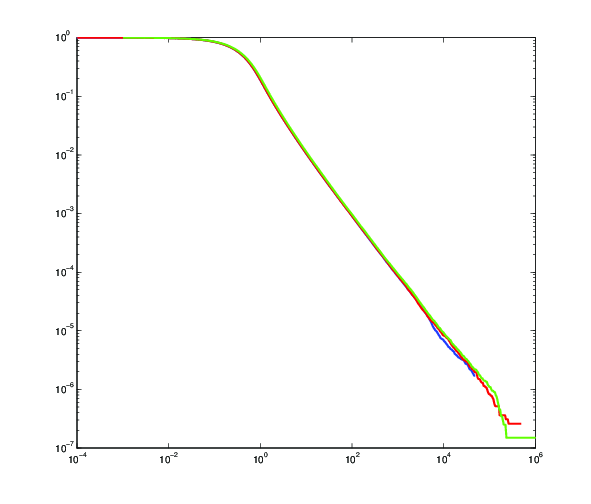}        
  \caption{The free path length distribution in the quasi crystal, for
  $\varepsilon=10^{-5}, 10^{-4}$ and $10^{-3}$. To the left the curves
represent the number of steps and on the right, the curves represent
the scaled time before a trajectory hits an obstacle.}
  \label{fig:sim3}
\end{figure}

Finally, Figure~\ref{fig:sim3} shows the path length distribution
$\phi_{\mathcal{F},\varepsilon} ([T,\infty[)$ for the quasi crystal
with three different values of $\varepsilon$. The graph on the left
shows distribution of the number of free steps ,
$k_{\mathcal{F},\varepsilon}$, and the graph on the right shows the
three curves expressed as a function of $T$, as in
Equation~\ref{eq:pathlenghts2}. That these curves almost coincide is a
strong indication that $\phi_{\mathcal{F},\varepsilon}[T,\infty[)$
converges to some distribution $\phi_{\mathcal{F}}[T,\infty[)$ as
$\varepsilon\rightarrow\infty$, just as in the periodic case.

\section{Conclusions}
\label{sec:conclusions}

The simulation results show that at least the one-dimensional discrete
time Lorentz gas in a quasi crystal behaves essentially as the
periodic Lorentz gas, and at least that the path length distribution
is not exponentially decaying, as in the random case.  The simulation
examples all refer to the specific case of a Fibonacci sequence based
quasicrystal. There are other constructions of onedimensional
quasicrystals, and in particular one can construct a whole family of
point distributions constructed as non-peridic sequences of two
different intervals (the book by Senechal~\cite{Senechal1995book}
presents  some, and give many references). In particular I wanted to
see whether the special choice of the golden ratio as ration between
long and short intervals would give a qualitatively different result
from other, say trancendent, ratios. But the qualitative behaviour
seems to be the same, and hence no simulation results are presented here.

However, the
simulations have been carried out in a much simplified model of the
gas, and therefore it is not obvious that a simulation of a real
(two-dimensional) Lorentz gas in a quasi crystal would give the same result.

Because the quasi crystals considered here are constructed as
projections of a regular lattice in higher dimension, it is possible
that the methods of Marklof and Strömbergsson referred to above would
also give results in this case. This possibility is mentioned
in~\cite{Marklof2010}, but as of now, no further results have been
published.

On the other hand, there are rather explicit results for a discrete
Schrödinger equation in a one-dimensional quasi crystal like the one
studied numerically here~\cite{Ostlundetal1983}.

\section*{Acknowledgments}
\label{sec:acknowledgments}
I would like to thank Andreas Strömbergsson for kindly providing
information on his and Marklof's work, and for pointing out several
relevant references. I would alslo like to thank Emanuele Caglioti and
Fran{\c cois Golse for many intersting discussions. The research has
been partially supported by the 
Swedish Research Council.

\bibliographystyle{plainnat}

\bibliography{./kinetikreferenser,./bernt_biblio}

\end{document}